\documentclass[conference]{IEEEtran}
\IEEEoverridecommandlockouts

\usepackage{cite}
\usepackage{amsmath,amssymb,amsfonts}
\usepackage{graphicx}
\usepackage{textcomp}
\usepackage{xcolor}
\usepackage{url}
\usepackage{float}
\usepackage{listings}

\usepackage{algorithm}
\usepackage{algpseudocode}

\begin{document}


\title{Electromagnetic Side-Channel Attack Resilience against PRESENT Lightweight Block Cipher\\
\thanks{*This work is supported by the research grants from the School of Computing, Edinburgh Napier University, UK. Any correspondence related to this article can be sent to \texttt{nilupulee.gunathilake@napier.ac.uk}}
}

\author{\IEEEauthorblockN{\textbf{Nilupulee A. Gunathilake, Ahmed Al-Dubai, William J. Buchanan and Owen Lo}}
\IEEEauthorblockA{Blockpass ID Lab, School of Computing, Edinburgh Napier University, United Kingdom}
}

\maketitle


\begin{abstract}

Lightweight cryptography is a novel diversion from conventional cryptography that targets internet-of-things (IoT) platform due to resource constraints. In comparison, it offers smaller cryptographic primitives such as shorter key sizes, block sizes and lesser energy drainage. The main focus can be seen in algorithm developments in this emerging subject. Thus, verification is carried out based upon theoretical (mathematical) proofs mostly. Among the few available side-channel analysis studies found in literature, the highest percentage is taken by power attacks. PRESENT is a promising lightweight block cipher to be included in IoT devices in the near future. Thus, the emphasis of this paper is on lightweight cryptology, and our investigation shows unavailability of a correlation electromagnetic analysis (CEMA) of it. Hence, in an effort to fill in this research gap, we opted to investigate the capabilities of CEMA against the PRESENT algorithm. This work aims to determine the probability of secret key leakage with a minimum number of electromagnetic (EM) waveforms possible. The process initially started from a simple EM analysis (SEMA) and gradually enhanced up to a CEMA. This paper presents our methodology in attack modelling, current results that indicate a probability of leaking seven bytes of the key and upcoming plans for optimisation. In addition, introductions to lightweight cryptanalysis and theories of EMA are also included.

\end{abstract}





\section{Introduction}


Internet-of-things (IoT) is a communication infrastructure that is being spread widely, increasing the number of connected devices exponentially. Estimates predict that there would be more than 200 billion connected devices by 2025 \cite{Gremban2018}. The ecosystem of IoT is constrained in terms of resource adaptability, because it is operated on low data rates (kbps), contains small onboard memories and is usually battery-powered. Nevertheless, its data flow is known to be dense, opaque and supposed to have low latency. Power consumption of these devices is greatly reduced in comparison with standard computing devices. Hence, green networking is an added advantage of IoT. An overall review of IoT is accessible in \cite{nilu_ICIA20} and \cite{nilu_WFToT19}. However, IoT is struggling to adopt adequate security features because conventional cryptography requires high processing capabilities, large capacities as well as faster data rates. As a result, a specific approach just targeting IoT data and privacy protection was introduced recently to build cryptographic methods in lightweight. Those techniques expect to offer shorter key lengths/initialisation vector (IV), smaller block sizes/internal state (IS) and lower memory requirements. A complete literature review about lightweight cryptology can be referred in \cite{nilu_CNSM20}.

In cryptology, cryptanalysis is vital to verify the strengths and weaknesses of proposed cryptographic algorithms. In cryptanalysis, non-cryptographic primitives such as internal power variations, external electromagnetic (EM) radiation, acoustic changes, data remanence on devices pose a substantial threat in securing a device. Therefore, side-channel analysis of recommended ciphers is a must in parallel with other types of cryptanalysis such as mathematical validations, use-case simulations and brute-force analysis. The basics of physical security phenomena are accessible in \cite{nilu_CRYPIS21}. Study \cite{SCAKlein} demonstrates that KLEIN is a side-channel resistant cipher regarding first-order attacks, but it may still be vulnerable to higher-order attacks. Recently, an option of re-keying which helps prevent side-channel attacks, has been introduced to lightweight cryptography as well \cite{ISAPLWC}. On the other hand, physical leakage analysis remains to be thoroughly researched. The majority of existing work belongs to power analysis (PA) \cite{SCA_LWC}. In this context, \cite{DPA_PRESENT1,DPA_PRESENT2} and \cite{OwenPRESENT} are about differential PA (DPA) and a correlation PA (CPA) of PRESENT respectively. A DPA of Simon and LED is available in \cite{SCA_LWC3} and a CPA of Fantomas, LBlock, Piccolo, PRINCE, Simon and Speck is accessible in \cite{SCA_LWC2}. However, other crucial characteristics such as EM emission, cache monitoring, optical changes, cold boot remain to be fully observed. \cite{DEMA_PRESENT} evaluates results against a differential EM analysis (DEMA) of PRESENT. \cite{EMA_prince} and \cite{twine} are about correlation EMA (CEMA) of PRINCE and Twine respectively. According to the available literature, a research outcome regarding CEMA of PRESENT by another research group is still unavailable.


\subsection{Our Contribution}


PRESENT is a promising block cipher recognised to be an alternative for Advanced Encryption Standard (AES) in lightweight applications. According to the developers of the cipher, it is more prone to side-channel and invasive hardware attacks \cite{PRESENT}. Thus, our contribution involves modelling a white-box, but non-invasive CEMA attack to evaluate the vulnerability of the PRESENT against its firmware robustness. This is still ongoing research, and this paper structures over:
\begin{itemize}
    \item An EMA classification and its relevant theories
    \item A description of our attack model implementation
    \item Our latest results and observations
    \item Discussion over the progress achieved so far
    \item Plans for optimisation and finalisation of the work
\end{itemize}


\section{Electromagnetic Side-Channel Analysis}


Electronic circuitries emit EM radiation as they operate. The radiated EM emanation can be detected using near-field (NF) EM compatibility (EMC) probes. According to Faraday’s law of induction, changes of magnetic flux in a magnetic field generate a voltage in the probe’s loop (equation \ref{eq:Faraday}). In EMA, excess EM radiation round a device resulted by an encryption is measured to observe if there is any correspondent relationship between secret information and EM field variations. However, the task is more difficult from the attacker’s perspective where prerequisite knowledge of the encryption key is unavailable. Although oscilloscope has been the typical device used to monitor and collect EM waveforms, software-defined radio (SDR) has become an interesting low-cost alternative nowadays.

\begin{equation} \label{eq:Faraday}
V = 2 \pi B A
\end{equation}

\textit{where,}

\textit{V - Voltage}

\textit{$\pi$ - The constant Pi, equal to 3.14159}

\textit{B - Average magnetic field}

\textit{A - Area perpendicular to the magnetic field}

If a possibility of any EM attack is indicated in preliminary studies, necessary countermeasures can be enabled in prior to manufacturing devices for commercial use, such as:
\begin{itemize}
    \item Proper EM shielding made of suitable materials, \textit{e.g., inclusion of Faraday cages}
    \item Addition of EM noise to hide or misguide the leakage
    \item Asynchronism of device clock correspondent to critical cryptographic functions
    \item Cryptographic operation obfuscating firmware application \cite{Sayakkara_2019}
    \item Randomisation of cryptographic function sequences and or lookup tables
    \item Use of pointers in data structures instead of values
\end{itemize}

There are several attack models used in EMA, known as simple EMA (SEMA), DEMA, CEMA and template EMA (TEMA). Since our work is based on SEMA and CEMA, those two types are briefly described under the following subsections. Despite the type, Hamming calculations are an essential procedure to obtain hypothesised values to compare with actual data \cite{CEMA_eq}. Hamming results indicate the maximum number of bit changes within the registers of the device. For obtaining the values, either Hamming distance (HD) or Hamming weight (HW) method is used. In this study, the HW (equation \ref{eq:HW}) has been used due to its higher efficiency. This counts all numbers of non-zero elements in a binary number at once, \textit{e.g., HW of 10110010 is 4}.

\begin{equation} \label{eq:HW}
E = a.HW(D)+b
\end{equation}

\textit{where,}
     
\textit{E - Hypothesised EM emission energy}
     
\textit{D - Intermediate value}
     
\textit{a - Gain}
     
\textit{b - Noise}

Performing EMA has been conducted in the time domain for a known period of time. On the contrary, new efforts were introduced in the frequency domain recently as an improved step. According to the literature, frequency domain work tends to avoid trace misalignment issues where time-domain results may be affected by frequently.


\subsection{Simple EMA (SEMA)}


This is simply a visual inspection of EM traces to identify its leakage points or encryption behaviour. Generally, the process may not involve breaking into secret data, but it might become possible to extract encryption keys by contemplating clock information as well as presumable HW changes of the device \cite{E_SCA_Sim}.


\subsection{Correlation EMA (CEMA)}


This is an efficient version of DEMA which processes several bits at a time where device details are not required. The computations focus on the correlation between a hypothesised intermediate value obtained via either the HD or the HW method and actual data captured in EM traces. The highest correlation of accurately aligned traces may indicate a possibility of a leakage point. Equation \ref{eq:correlation} is used to calculate correlation coefficient for the task.

\begin{equation} \label{eq:correlation}
\rho = \frac{Cov(X,Y)}{\sigma_X \sigma_Y}
\end{equation}

\textit{where,}

\textit{$\rho$ - Pearson correlation coefficient}

Cov(X,Y) - Covariance between X and Y

\textit{$\sigma_X$ - Standard deviation of X}

\textit{$\sigma_Y$ - Standard deviation of Y}


\section{Attack Modelling}


\subsection{PRESENT Block Cipher}

\begin{figure}[!htbp]
        \center{\includegraphics[width=0.35\textwidth]
        {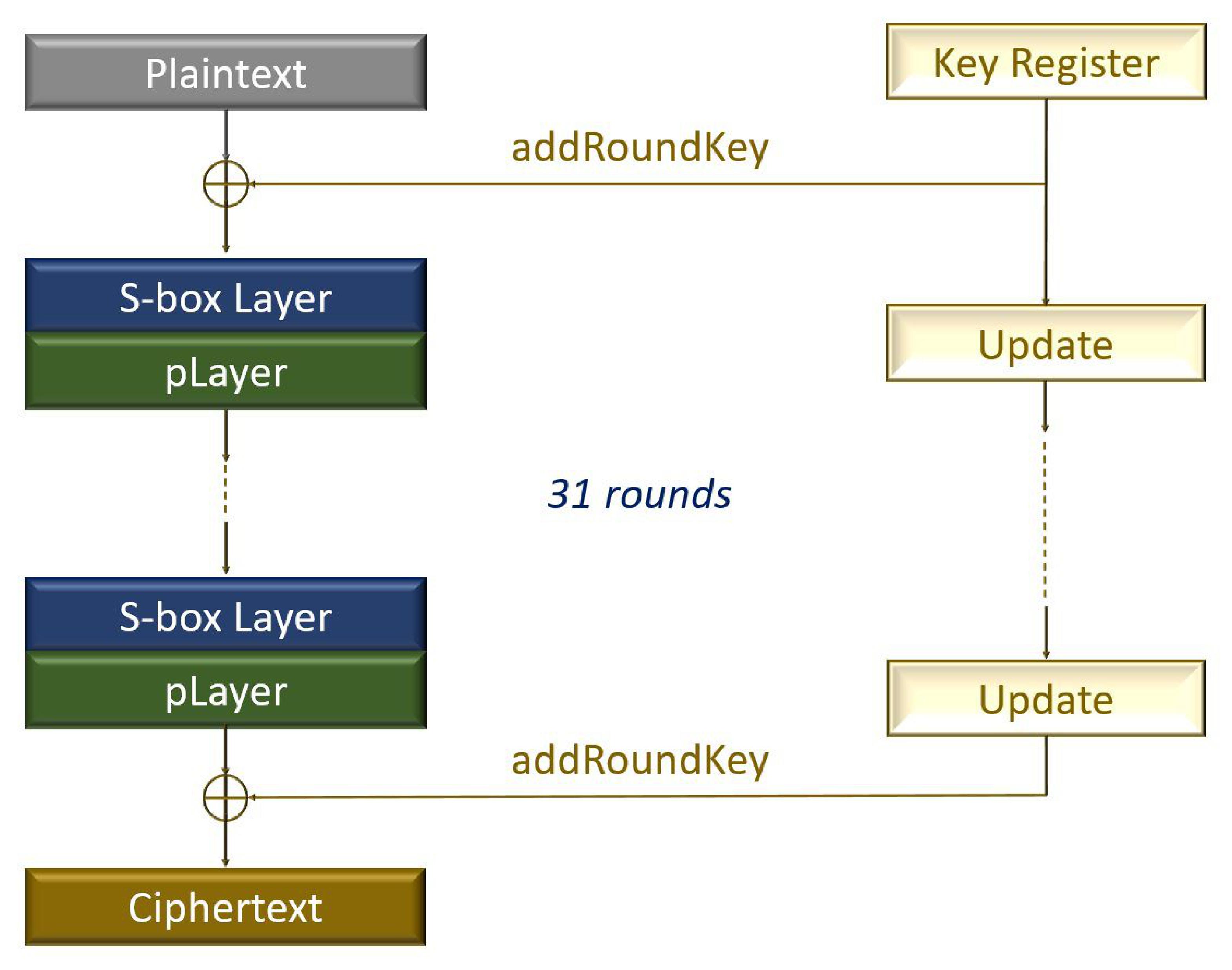}}
        \caption{\label{fig:PRESENT}PRESENT encryption process}
\end{figure}

PRESENT is a block cipher introduced by the authors of \cite{PRESENT} in 2007. It is recognised to be an ultra-lightweight\footnote{Ultra-lightweight cryptography targets specific areas of algorithms for selective hardware types and or selected cipher sections} cipher that has been approved by the ISO/IET \cite{ISOIET29167}. In addition, the NIST has mentioned it under lightweight block cipher listing in their NISTIR 8114 report \cite{NISTIR8114}. Its architecture is a substitution-permutation network (SPN), and the block size is 64-bit. Although there are two versions with a 80-bit key and a 128-bit key, the 80-bit one is recommended for lighter weight encryption. The energy consumption is around 5$\mu$W over 32 clock cycles. It computes through 31 rounds as in Fig. \ref{fig:PRESENT}. This cipher aims hardware optimisation owning small footprints of 1570 gate equivalent (GE) for the 80-bit version and 1886 GE for the 128-bit version. The substitution box (S-box) is a 4-bit to 4-bit which results in 28 GE. The numerical mapping of it as in Table \ref{table:sbox}.

\begin{table*}[!htbp]
\centering
\caption{S-box mapping of PRESENT block cipher}
\label{table:sbox}
\begin{tabular}{|c|c|c|c|c|c|c|c|c|c|c|c|c|c|c|c|c|}
\hline
x & 0 & 1 & 2 & 3 & 4 & 5 & 6 & 7 & 8 & 9 & A & B & C & D & E & F \\
\hline
S(x) & C & 5 & 6 &	B &	9 &	0 &	A &	D &	3 &	E &	F &	8 &	4 &	7 &	1 &	2 \\
\hline
\end{tabular}
\end{table*}


\subsection{Methodology}


Initially, a SEMA of data distribution differences was performed for both encryption and non-encryption statuses. The primary resources used here are an oscilloscope (Keysight InfiniiVision MSOX4101A) with 5GSa/s (5 billion samples per second) and NF EMC probes (TekBox H20, H10 and H5) with a 9kHz to 6GHz frequency range. The signals were amplified using a 40dB wide-band amplifier (TekBox TBWA2) before being fetched to the oscilloscope. The encryption was run on an Arduino UNO board. Two probe positions were examined that are in parallel and perpendicular to the chip. MATLAB(r) 2020b software was used for postprocessing data. Regarding the choices of functionalities of the PRESENT:
\begin{itemize}
    \item 80-bit version is chosen
    \item The first round of the encryption is considered
    \item S-box was targeted due to its non-linearity. Thus, it would be easier to identify impacts on waveforms
    \item The encryption key used is AC DE FB 21 F9 23 75 C0 E6 as same as in \cite{OwenPRESENT}
\end{itemize}

A trigger signal was used to locate the S-box operational area of the waveform by connecting the LED port of the Arduino UNO board to a separate channel of the oscilloscope. Consequently, the usable sampling rate for EM traces was reduced to 2.5GSa/s. The Arduino IDE code used for PRESENT encryption was derived from \cite{arduinocode}, and its accuracy was verified using test vectors given in \cite{PRESENT}. In contrast, MATLAB codes created for trace collection, reconstruction and attack performance were validated using known test data values. Some precautions were taken to enhance the performance by reducing possible system noises and ambient EM interferences. For that:
\begin{itemize}
    \item A resultant averaged waveform for five encryption cycles was taken per ciphertext
    \item The setup was lightly covered using Faraday fabrics (low-cost alternative instead of expensive Faraday cages)
    \item The computer was operated in flight mode
    \item New frequency components generated as a result of the encryption were filtered
\end{itemize}

At our current stage, 256 waveforms were collected for 256 different plaintext values, each byte of the plaintext value incrementing from 0x00 to 0xFF in hexadecimal. Firstly, the encryption code was set for the first round, compiled and uploaded on the board. The output of the previous AddRoundKey has to be taken into attention when the S-box function is defined in encryption (Arduino IDE) as well as postprocessing of actual data (MATLAB). Regarding the CEMA, hypothesised calculations for ciphertexts were obtained in MATLAB considering each key byte value from 00 to FF for each plaintext used during encryption. Then, the HW of the ciphertexts were gained as follows.

\begin{algorithm}
\label{alg:HW}
\caption{HW calculation of the ciphertexts} 
\begin{algorithmic}
    \For{$k=0,1,2,\ldots,255$}
        \For {$p=0,1,2,\ldots,255$}
	        \State Output of AddRound key step set input to S-box
		    \State Look up the S-box value
		    \State Calculate HW and save
	    \EndFor
    \EndFor
\end{algorithmic}
\end{algorithm}

Next, the highest correlation coefficient ($\rho$) values between the HW results and actual data points were calculated per plaintext. Using a graph of the data points versus $\rho$, correspondence key values for the highest correlation points were checked. Apart from just the highest correlation branches, the key-value distribution over the graph was analysed to identify potential leakage areas. The pseudo-code for calculation correlation values is shown below.

\begin{algorithm}
\caption{Correlation coefficient calculation} 
\begin{algorithmic}
    \For{$k=1,2,\ldots,last~data~point~of~waveform$}
	    \State Calculate $\rho$ between arrays of actual data and HW
	    \If{$empty$} 
		    \State Save key value and its $\rho$ value
		\Else
		\If{$\rho\geq previous$} 
		    \State Overwrite key value and update $\rho$ value
		\EndIf
		\EndIf
	\State Plot graph of data points vs. $\rho$
	\EndFor
\end{algorithmic}
\end{algorithm}


\section{Results and Observations}


\subsection{SEMA}


The parallel position data was often noisy, and it did not reflect any substantial difference between encryption and non-encryption statuses. Regarding the perpendicular position of all magnetic probes, appearing of new frequency elements could be observed at 11.25MHz, 22.5MHz, 45.08MHz, 56.33MHz, 78.83MHz, 90.08MHz and 112.66MHz. Among those, the 45.08MHz component has the highest amplitude and the 56.33MHz owns the second highest. However, increased amplitude of already existing elements during the non-encryption status could be seen at 33.75MHz, 50.62MHz, 67.58MHz, 135.16MHz, 151.87 and 168.98MHz. Histogram plots revealed a slight voltage increase during the encryption.


\subsection{CEMA}

\begin{figure}[!htbp]
        \center{\includegraphics[width=0.5\textwidth]
        {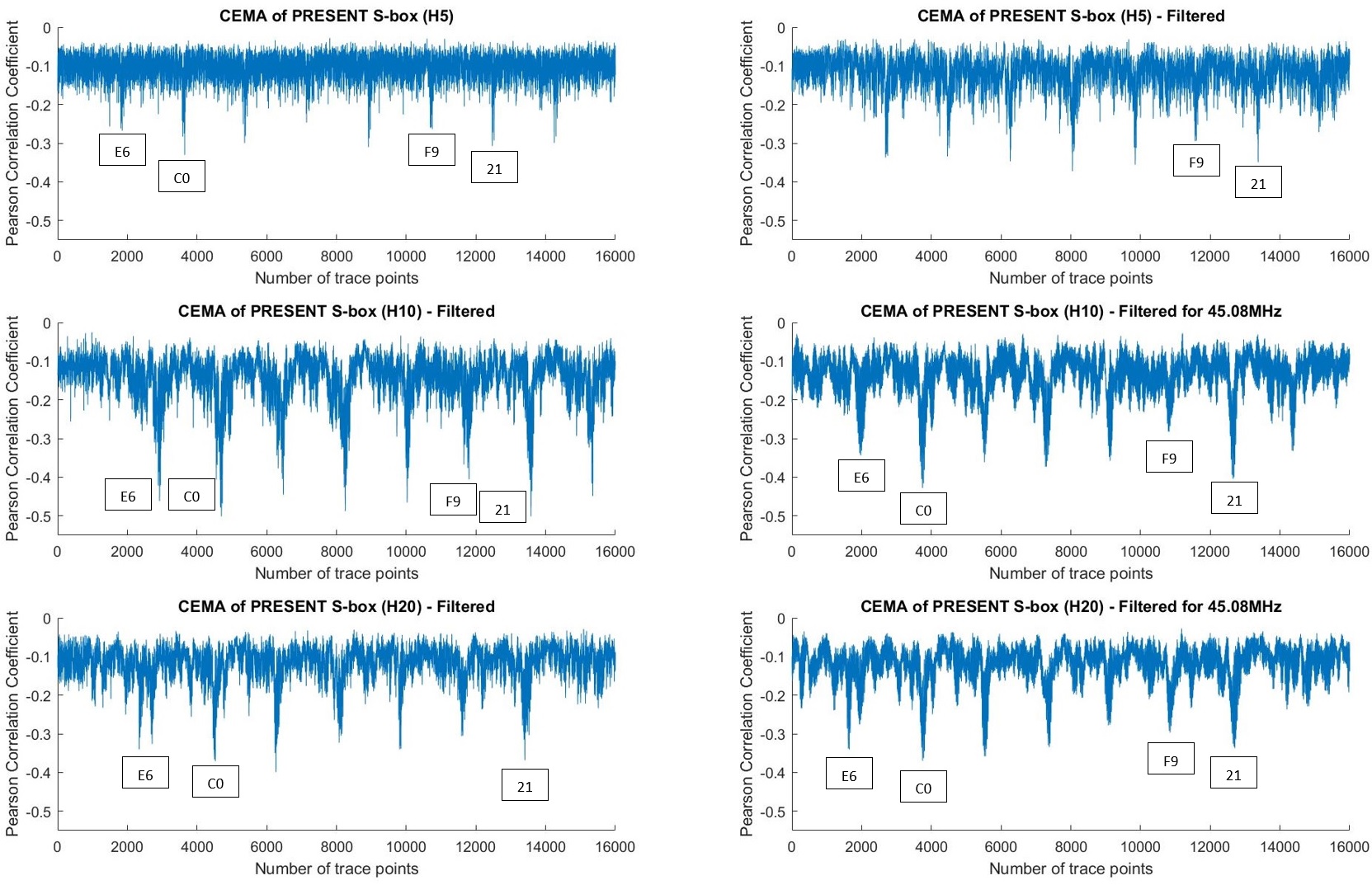}}
        \caption{\label{fig:cema}Result comparison of the CEMA of PRESENT}
\end{figure}

Even though around eight noticeable trough areas appeared in the correlation graphs similar to the power analysis results in \cite{OwenPRESENT}, a sharpened shape could be gained after filter application. At this phase, bandpass filters were tried regarding all new elements together and the first two highest components individually. The 45.08MHz element illustrated better shapes, and the 56.33MHz one did not show any promising pattern. Correlation graph comparison for some sample data as in Fig. \ref{fig:cema}. Most of the time, the lowest point of the troughs did not reveal the exact key byte, but the correct key leakage was able to be found somewhere in the lower part of the relevant trough, up to seven bytes which are FB, 21, F9, 23, 75, C0 and E6. A summary of key byte indication as in Table \ref{table:cema}. In addition, the same method was run on the non-encryption data in order to verify that the notable troughs are due to the encryption impact. No significant correlation difference was there in non-encryption data.

\begin{table*}[!htbp]
\centering
\caption{Key byte leakage probabilities}
\label{table:cema}
\begin{tabular}{|c|c|c|c|c|c|c|c|c|}
\hline
\multicolumn{9}{|c|}{Probability of Leakage}\\
\hline
Key Byte & \textbf{FB} & \textbf{21} & \textbf{F9} & \textbf{23} & \textbf{43} & \textbf{75} & \textbf{C0} & \textbf{E6} \\
\hline
H5 & 0 & 33\% & 60\% & 6.67\% & 0 & 0 & 20\% & 46.67\% \\
\hline
H10 & 13.33\% & 53.33\% & 66.67\% & 0 & 0 & 6.67\% & 20\% & 80\% \\
\hline
H20 & 6.67\% & 53.33\% & 46.67\% & 6.67\% & 0 & 13.33\% & 20\% & 40\% \\
\hline
\multicolumn{9}{|c|}{Probability of Leakage at a Time}\\
\hline
H5 & \multicolumn{2}{|c|}{Four bytes: 6.67\%} & \multicolumn{2}{|c|}{Three bytes: 13.33\%} & \multicolumn{2}{|c|}{Two bytes: 46.67\%} & \multicolumn{2}{|c|}{One byte: 13.33\%}\\
\hline
H10 & \multicolumn{2}{|c|}{Four bytes: 20\%} & \multicolumn{2}{|c|}{Three bytes: 40\%} & \multicolumn{2}{|c|}{Two bytes: 13.33\%} & \multicolumn{2}{|c|}{One byte: 13.33\%}\\
\hline
H20 & \multicolumn{2}{|c|}{Four bytes: 6.67\%} & \multicolumn{2}{|c|}{Three bytes: 26.67\%} & \multicolumn{2}{|c|}{Two bytes: 33.33\%} & \multicolumn{2}{|c|}{One byte: 20\%}\\
\hline
\end{tabular}
\end{table*}


\section{Discussion}


The summary of the results confirms that the encryption has affected the EM emission of the device, and it illustrates a high probability of leaking at least two bytes at once up to seven bytes. Not having significant troughs regarding non-encryption data verifies the above fact further. In contrast, the potential of the first (E6), the sixth (F9) and the seventh (21) byte leakage is greater as the probability exceeds 50\%. The H5 (5mm diameter) probe revealed four bytes, the H10 (10mm diameter) did six bytes and the H20 (20mm diameter) leaked seven bytes. The H5 was able to gain the maximum number of bytes at once (four bytes) without even needing to being filtered, but it could not leak all seven bytes. What is more, the H20 was able to offer all seven bytes, but in individual attempts. This may be due to the fact that probes with larger loops are more sensitive, but have lower frequency resolution. However, extracting the exact bytes correctly is quite challenging because external noise interference cannot be avoided completely in EMA. Nevertheless, changes of frequency selection and filter orders in bandpass filtering caused the location change of indicated bytes most of the time.  Therefore, a choice of the most suitable filter type along with its order, as well as the position must be made for optimised results.

According to Fig. 7 of \cite{DEMA_PRESENT}, the accuracy of leakage increases when the number of waveforms are increased. However, the possible maximum number of waveform collections may depend on the design of the attack model. \cite{Sayakkara_2019} mentions that filtering frequencies closer to harmonics of the device clock frequency may increase exactness. Hence, our further steps to enhance the performance by:
\begin{itemize}
    \item Analysing all affected frequencies individually (newly appearing and amplitude changed ones) and clock frequency harmonics in filtering
    \item Verifying over the results via the most suitable probe type, filter type, order and frequency range
    \item Increasing the number of waveforms if possible (up to 2048)
    \item Increasing the sampling rate up to 5GSa/s if possible
    \item Calculating the success rate for different key values
\end{itemize}

At least one-byte leakage of the key reduces processing time considerably in brute force analysis to derive the rest of the bytes. On the other hand, it is extremely difficult for an attacker to locate the functional area when the encryption runs in all 31 rounds with all dependable steps such as addRoundKey, S-box and Player in a black-box environment. This study further verifies that the optimum position of Arduino UNO for EM attack is perpendicular between +5V and GND pins as in \cite{peter2019}.


\section{Conclusions}


IoT devices are struggling to have sufficient security due to their constraints regarding resource adaptability. Also, this smart technology introduces smarter threats and hazards. Thus, the integration of sufficient security mechanisms is challenging. Consequently, lightweight cryptography comes to the rescue. However, lightweight ciphers are emerging and need careful verification over proper analyses before their use in commercial applications. Side-channel attacks are of the utmost importance in physical security because it is a different scenario from algorithmic strength. Our research focuses on EM side-channel resilience against PRESENT lightweight block cipher. The work initially started from a SEMA and was then enhanced up to a CEMA. There is no other CEMA of the PRESENT existing in the literature. The current results illustrate eight leakage locations with a probability of encryption key leakage up to seven bytes out of ten. This work still continues towards optimisation.



\bibliographystyle{IEEEtran}
\bibliography{bib}

\end{document}